\documentstyle[aps,prl]{revtex}
\draft

\begin{document}
\twocolumn

\title{
Relaxation, the Boltzmann-Jeans Conjecture and Chaos}

\author{Naoko NAKAGAWA}
\address{
Department of Mathematical Science, Faculty of Science,
Ibaraki University,
Mito, Ibaraki 310-8512, Japan}

\author{Kunihiko KANEKO}
\address{Department of Pure and Applied Sciences,
College of Arts and Sciences,
University of Tokyo,
Tokyo 153, Japan}

\maketitle

\begin{abstract}

Slow (logarithmic) relaxation from a highly excited state is studied in a
Hamiltonian system with many degrees of freedom.  The relaxation
time is shown to increase as the exponential of the square root of the
energy of excitation,
in agreement with the Boltzmann-Jeans conjecture, while
it is found to be inversely proportional to residual
Kolmogorov-Sinai entropy, introduced in this Letter.
The increase of the thermodynamic entropy through this relaxation
process is found to be proportional to this quantity.

\end{abstract}


Study of the relaxation from highly excited states is important
not only in physico-chemical systems but also in biological systems.  
It has been reported that excitations of some protein molecules
 are maintained over anomalously long
time spans \cite{Ishijima_Yanagida}.
Such behavior is relevant to enzymatic reactions, 
and in particular  to their biological functions,
because these molecules must maintain their excited state (brought about,
say, by ATP) to be able to proceed from one process to the next.

Typically, a Hamiltonian system with a sufficiently
large number of degrees of freedom relaxes to equilibrium.
This relaxation process, however, is not necessarily fast.
For instance, some Hamiltonian systems can exhibit
logarithmically slow relaxation from excited states, 
called Boltzmann-Jeans conjecture (BJC),
which was first noted by Boltzmann\cite{Boltzmann},
explored by Jeans\cite{Jeans}, Landau and Teller\cite{Landau_Teller},
and then has been studied in terms of nonlinear dynamics
\cite{Benettin_Galgani}.
In the BJ conjecture, the relaxation to equilibrium is
required, but existence of chaos is not explicitly assumed.
On the other hand, irreversible relaxation  is often studied in
relationship with chaos.
In the present paper we study slow relaxation of a type
in agreement with the BJC, by using a Hamiltonian 
dynamical system with a large number of degrees of freedom,
and explore its possible relationship with chaos.


We consider dynamics with the Hamiltonian
\begin{equation}
H=K+V=\sum\limits_{i=1}^N {p_i^2\over 2}+\sum_{i,j=1}^N
V(\theta_i,\theta_j)
\label{eqn:Hamiltonian}
\end{equation}
describing a set of pendula,
where $p_i$ is the momentum of the $i$-th pendulum.
The potential $V$ is chosen so that
each pair of pendula interacts only through their phase difference
\cite{KK,Antoni_Ruffo}:
\begin{equation}
V(\theta_i,\theta_j)  =
{1 \over {2(2\pi)^2N}}\{1-\cos(2\pi(\theta_i-\theta_j))\}.
\end{equation}
Hence, the evolution equations for the momentum $p_i$ and the phase
$\theta_i$ are given by
\begin{equation}
\dot p_j = {1\over{2\pi N}}\sum_{i=1}^N \sin(2\pi(\theta_i-\theta_j)),
~~~\dot \theta_j = p_j.
\label{eqn:seijunQ}
\end{equation}
This form of the interaction is chosen so that there is an attractive force 
tending to align the phases of all the pendula.
The thermodynamic properties of this model in the $N\rightarrow\infty$,
as well as the finite-size effect,
are investigated in Ref.\cite{Antoni_Ruffo,Latora_Ruffo}.
Here we mainly discuss the case with $N=10$,
which is sufficiently large to exhibit the thermodynamic properties, and
the results to be discussed are valid for a larger system.
The temperature of the system can be defined as $T\equiv \langle 2K\rangle/N$,
which is a monotonically increasing function of the total energy $E$.

As studied in Refs.\cite{KK,Antoni_Ruffo,Latora_Ruffo,Yamaguchi},
each pendulum in this system exhibits small-amplitude vibration
when the total energy is small,  while, as the energy is increased,
 some pendula begin to display rotational behavior over many cycles.
The relaxation of the rapidly rotating pendula into a vibrating assembly 
is rather slow,
since their average interaction with the assembly
almost cancels out over the slow time scale of the assembly,
due to the rapid rotation.
Here we concentrate on such slow relaxation
of the rotational mode to the vibrational modes. 

The slow relaxation from a highly excited state is investigated 
systematically in the situation that the excited state is prepared 
by applying an  instantaneous kick to a certain pendulum in the system.
An example of the relaxation process following this kick is
depicted in Fig.\ref{fig:eng_str}, where it is seen that
the kicked pendulum continues to rotate in isolation, maintaining
a large energy.  The relaxation time $t_R$ is defined
as the interval of time from the kick to the point of which 
kinetic energy of the kicked pendulum first becomes smaller than
$K/N$. 

The relaxation process depends on the energy $E_0$ imported by the kick
and also on the state of the other $N-1$ pendula.
Hereafter we call this assembly the `bulk'.
The bulk state is parameterized by the total energy $E$
(or the temperature $T$)
of the system before the kick \cite{note1}.

The mean relaxation time $\langle t_R \rangle$ over an ensemble
of initial conditions was computed by fixing $E$ and varying $E_0$.  
As shown in Fig.\ref{fig:exponential},
the relaxation time increases exponentially with the kicked energy $E_0$
\cite{Nakagawa_Kaneko} as
\begin{equation}
\langle t_R \rangle \propto \exp(\alpha \sqrt{E_0}),
\label{eqn:exponential}
\end{equation}
for sufficiently large values of $E_0$.
Here $\langle\cdot\rangle$ represents an ensemble average
and $\alpha$ is a constant.
When the value of $E_0$ is fixed, 
$\langle t_R\rangle$ decreases as a function of $E$,
as is seen in the figure.

The above exponential form of the relaxation agrees with that given
by the BJC,
which describes the relaxation time for a system consisting of 
two parts with very different frequencies 
as an exponential of the ratio of the two frequencies.
Assuming the separation of time scales of the kicked pendulum and
the bulk pendula,
this exponential growth can be obtained through the Fourier analysis
of the interaction term, as shown by Landau and Teller \cite{Landau_Teller}
(see also \cite{Benettin_Galgani}).

In this argument, chaotic dynamics are not explicitly included, 
while chaos is often relevant to the irreversible relaxation
to equilibrium.
Actually, we have observed  the exponential law (\ref{eqn:exponential})
in numerical studies clearly, 
as long as the Hamiltonian dynamics for the given energy $E$ 
uniformly exhibits chaotic behavior without
remnants of KAM tori.  In this regime, 
$\langle t_R \rangle$ decreases with the energy $E$,
as shown in the inset of Fig.\ref{fig:exponential}.
On the other hand, in the regime of the small values of $E$,
the relaxation process highly depends on the initial condition 
and it is difficult to define  average relaxation time.
The hindrance of relaxation is associated with the weakness of chaos, 
since the relaxation of the rotating pendulum stops when the dynamics 
of the bulk is trapped to the vicinity of remnants of tori.  Now,
we study a possible relationship between 
the relaxation time and the chaotic dynamics of the bulk.

First let us consider the nature of the dynamics in the
limit $E_0\rightarrow\infty$.
In this case, the kicked pendulum is completely free from the bulk of
vibrating pendula, whose collective motion is simply that determined
by the Hamiltonian of its $(N-1)$ pendula.
At a finite value of $E_0$, the dynamics of the bulk deviate from those in
the $E_0\rightarrow \infty$ limit, 
due to the interaction with the kicked pendulum.
In order to quantify such deviation,
we choose bulk states with energies distributed around a given $E$,
and compute the Lyapunov spectrum $\lambda_i(E_0)$ for various values of $E_0$.
Note that, in the limit $E_0\rightarrow \infty$, $\lambda_{N-2}=0$
corresponding to the free rotation of the kicked pendulum, while
$\lambda_{N-1}=\lambda_N=0$ always holds
due to the conservation of the total momentum and energy.

Since the kinetic energy $K_r$ of the kicked pendulum fluctuates and relaxes,
$\lambda_i(E_0)$ is computed
in the case that $K_r$ satisfies
$|E_0-K_r|<\delta $, with a small constant $\delta$.  
Since the relaxation of $K_r$ is very slow,
this condition is satisfied over a time sufficiently
long to compute the local exponents.  
In order to insure the  convergence of the Lyapunov exponents, 
we take an ensemble average of such local exponents \cite{note2}.
(The existence of such time scale for the convergence is guaranteed
by the slow relaxation, and indeed is confirmed numerically.)

The values of the computed Lyapunov exponents increase as $E_0$ decreases.
The increased part of the Lyapunov exponents from those at
$E_0\rightarrow \infty$ limit characterizes the
amplification of the chaotic instability due to the interaction between
the kicked pendulum and the bulk.  
With this in mind, we define the residual Lyapunov spectra by
\begin{equation}
\Delta\lambda_i(E_0)\equiv \lambda_i(E_0)-\lim_{E_0\rightarrow
\infty}\lambda_i(E_0).
\end{equation}
>From detailed numerical experiments,
we find that the two scaling relationships;
\begin{eqnarray}
&&\lambda_{N-2}(E_0)=\Delta\lambda_{N-2}(E_0) \propto \exp(-\beta_1 \sqrt{E_0}),
\label{eqn:N-2-exponential}\\
&&\Delta h(E_0) \equiv \sum_{i=1}^{N-3}\Delta\lambda_{i} \propto
\exp(-\beta_2\sqrt{E_0})
\label{eqn:KS-exponential}
\end{eqnarray}
hold for sufficiently large values of $E_0$.
Here $\beta_1$ is smaller than $\beta_2$, as shown in Fig.\ref{fig:KS}(a).

Since $\lambda_{N-2}=0$ in the limit $E_0 \rightarrow \infty$, 
the scaling of $\lambda_{N-2}$ represents the decrease 
in the degree of chaos of the kicked pendulum.
The sum of the remaining positive Lyapunov exponents, 
then, is expected to correspond to the Kolmogorov-Sinai (KS) entropy 
of the bulk, and 
the increase in the degree of chaos of the bulk due to the interaction with
the kicked pendulum is given by $\Delta h$.
(Note that the increase of Lyapunov exponents is not due to
the increase of the average velocity in the bulk.
Indeed, we have numerically confirmed that the temperature of the bulk for finite $E_0$
is not shifted from that in the limit $E_0\rightarrow \infty$.)

Since $\Delta h$ is associated with the dynamics of the bulk
and how it is affected by the energy absorbed from the kicked pendulum, 
we expect that it is related to the relaxation time $\langle t_R\rangle$
\cite{note3}.
Indeed $\alpha$ in Eq.(\ref{eqn:exponential}) and 
$\beta_2$ in Eq.(\ref{eqn:KS-exponential}) appears to be equal
from our numerical results.  
The relationship between $\langle t_R\rangle$ and $\Delta h$ 
is plotted in  Fig.\ref{fig:KS}(b), which
supports the relationship
\begin{equation}
\Delta h \propto \langle t_R \rangle^{-1}.
\label{eqn:KS_tR}
\end{equation}
Thus, we find that the degree of chaos is closely related to
the relaxation time scale.

 The relationship between the relaxation and the residual
KS entropy $\Delta h$ can be roughly explained as follows:
In a Hamiltonian system, for every orbit, there exists a time-reversed orbit, 
obtained by changing the sign of all $p_i$.  
However, because of the interaction with the kicked element, 
to obtain the time-reversed orbit of the bulk, 
a slight modification of each pendulum,
in addition to the simple reversed of each of their momenta, 
$p_i\rightarrow-p_i$, is needed.
Since the KS entropy $h$ is a measure of the time rate
of loss of the information concerning the initial conditions 
as an orbit evolves,
the measure of orbits
that can be considered the reversed orbit of some given orbit over a time $t$
decreases with $\exp(-ht)$.  
Therefore, the time scale for the irreversible relaxation
is proportional to the inverse of the KS entropy difference between
the original and the reversed orbits.
Hence, the relationship Eq.(\ref{eqn:KS_tR})
between the time scale for the relaxation and the residual
KS entropy is reasonable.

Next, we study dynamics on the long time scale of the order 
of $\langle t_R \rangle $.
To compare the relaxation from various kicked energies $E_0$, it is
convenient to use the scaled time $\tau =t/\langle t_R \rangle$,
where $ \langle t_R\rangle$ is a function of $E_0$.
We note that the time scale of relaxation 
decreases as the energy of the kicked pendulum, $K_r$, decreases.
Then the above definition of $\tau$ implies 
$\tau= t/t_R(\tilde K_r)$ along the relaxation, where
$\tilde K_r$ is the coarse-grained energy of the kicked pendulum 
over a long time scale and $t_R \propto \exp(\alpha \sqrt{\tilde K_r})$.

Now, we derive an equation describing the relaxation 
of $\tilde K_r$ on the time scale of $\tau$.
We have computed the difference $\delta K_r$
over the time steps $\delta \tau$ for many samples of orbits starting 
from a given $K_r$.
As shown in Fig.\ref{fig:delta}, the data are satisfactory fit by
\begin{math}
\frac{\langle \delta K_r\rangle}{\delta \tau}= -C,
\end{math}
with a constant $C$ ($\simeq 0.035$) for various values of $K_r$ and $E$.  
Then the coarse-grained equation for the relaxation 
is obtained as
\begin{math}
{dK_r}/{d \tau}= -C.
\end{math}
Noting that the interaction energy between the bulk 
and the kicked pendulum is tiny and recalling the conservation of energy, 
we get
\begin{math}
{d E}/{d \tau}= C.     
\end{math}
Thus, with regard to the relaxation process,
the difference in the details of the dynamical properties of the system
under different conditions are eliminated 
by considering the scaled time $\tau$,
while the dynamics and properties of the bulk seem to be strongly 
dependent on the value of $E$.
>From the above equation, for the original time variable $t$, 
$K_r$ decays in the logarithmic time scale.

Now it is possible to consider the entropy increase of the bulk 
occurring during the relaxation.
Since we have assumed that the slow change of the
bulk part can be described by the slow change of thermodynamic quantities,
the increase of the entropy of the bulk is estimated as
\begin{equation}
\frac{dS}{d\tau}= \frac{dS}{dE} \frac{dE}{d\tau}=\frac{C}{T},
\label{eqn:S}
\end{equation}
where $T$ is the temperature of the bulk.
This relation is obtained by noting that the process here is sufficiently
slow to define these thermodynamic quantities on a coarse-grained time scale.

Since $\tau =t/t_R\propto \Delta h~t$, by Eq.(\ref{eqn:KS_tR}), 
the relation
\begin{equation}
\frac{dS }{dt} \propto \Delta h
\label{eqn:S2}
\end{equation}
for  the relaxation process is obtained from Eq.(\ref{eqn:S}).
The proportionality in Eq. (\ref{eqn:S2}) relates two kinds of 
directed motion towards equilibrium:
$\Delta h$ corresponds to the amplification of chaos
in the equi-energy surface of the bulk,
whereas $dS/dt$ represents the growth of this equi-energy surface
towards equilibrium.
It should also be noted that this relationship holds
not near the equilibrium state of the total system but for the nonlinear
relaxation process from a highly excited state.

The relationship between chaotic dynamics and irreversibility has been
extensively studied \cite{note3,Dzugutov_Vulpiani,Sasa_Komatsu}.  
A relationship between the thermodynamic entropy and
the irreversible information loss was proposed by Sasa and Komatsu 
in the case that an external operation causes a transition from one
equilibrium state to another. 
We believe that our relation between the  residual KS entropy 
and the thermodynamic entropy
is related to it, although ours applies to the spontaneous relaxation.

To sum up, we have obtained the following relationship between 
the relaxation time, $t_R$, and the energy supplied by an external kick, $E_0$:
$t_R \propto \exp(\alpha \sqrt{E_0})$. 
Then, 
this relaxation time $t_R$ is found to be 
proportional to the inverse of the residual KS entropy, 
that is, the difference between the KS entropy of the bulk
interacting with the kicked element and that of the isolated
bulk.  By rescaling time by $t_R$,
the excited energy is found to relax at a rate
that is independent of the energy of the  kicked element and the bulk.
Finally, we find evidence that the rate of entropy generation 
is proportional to the residual KS entropy.

Although these three relationships were found through numerical study of
the Hamiltonian (\ref{eqn:Hamiltonian}),
we expect that they hold generally for relaxation processes
in Hamiltonian systems 
under the condition that in the limit of high excitation, the
interaction between the excited elements and the unexcited bulk vanishes.
Indeed, we have numerically found that Eqs.(\ref{eqn:exponential}) and
(\ref{eqn:KS_tR}) are valid also for a coupled pendulum model 
with the nearest-neighbor coupling
on a square lattice.
It is important to examine
the universality of the three relationships we have found, 
and also to study their relevance in regard to intra-molecular 
relaxation processes, including
application to the energy transduction carried out by
biological molecular motors.  

The authors are grateful to
S. Sasa and T. S. Komatsu for stimulating discussions.  
This research was supported by Grants-in-Aids for Scientific 
Research from JSPS and the REIMEI Research Resources of JAERI.

\begin{figure}
\caption{Time series of the momenta $p_i$ ($1\le i \le N$) in the
process of energy relaxation after an instantaneous kick at $t=0$.
Here, $E/N=0.0205$, $N=10$ and $E_0=0.35$.
A is the time series of the kicked pendulum.
}
\label{fig:eng_str}
\end{figure}

\begin{figure}
\caption{
Ensemble average of the relaxation time $\langle t_R\rangle$ as a function 
of the square root of the injected energy $E_0$.
Here, ($\triangle$) $E/N=0.0205$ ($T=0.019$), ($\bullet$) $E/N=0.0118$ ($T=0.008$).
The dotted line represents $g(E_0)\equiv \exp(15\sqrt{2E_0})$.
In the inset, in order to see the close up temperature dependence of 
$\langle t_R\rangle$, $\langle t_R\rangle/g(E_0)$ is plotted 
as a function of $T$ for four values of $E_0$:
$E_0=0.35$ $(+)$,
$0.4$ $(\circ)$,
$0.45$ $(\times)$ and
$0.5$ $(\Box)$.
The data are fit by the function $13\exp(-135T)$ for small $T$.
}
\label{fig:exponential}
\end{figure}

\begin{figure}
\caption{
(a) Dependence of $\lambda_{N-2}$ $(\circ)$ and $\Delta h$ $(\Box)$ on $E_0$
for $E/N=0.0205$ $(T=0.019)$.
The data are fit by the lines 
$\Delta \lambda_{N-2} \propto \exp(4.1\sqrt{2E_0})$
 and $\Delta h\propto \exp(15\sqrt{2E_0})$.
(b) $\langle t_R\rangle$ as a function of $\Delta h$ 
for $E/N=0.0205$ $(T=0.019)$.
To calculate the Lyapunov Spectrum $\lambda_i(E_0)$, we adopt the criterion
$|E_0-K_r|<\delta$, with $\delta=0.00375$.
The calculated $\lambda_i(E_0)$ is found to converge for small $\delta$.
}
\label{fig:KS}
\end{figure}

\begin{figure}
\caption{
$\langle \delta K_r\rangle/\delta \tau$
for $\delta \tau=0.01$ $(\circ)$ and $0.001$ $(\Box)$.
upper: dependence on the temperature $T$.
lower: dependence on the energy $K_r$ of the kicked pendulum.
The data suggest that $\langle \delta K_r\rangle/\delta \tau$
is independent of $\delta \tau$, $K_r$ and $T$.
}
\label{fig:delta}
\end{figure}


\begin{thebibliography}{999}
\bibitem{note1}
Although the energy of the bulk increases with relaxation,
this parameterization is appropriate, at least in the
lowest order approximation for large $E_0$,
since 
the bulk and the kicked pendulum spend most of the relaxation time
around the state it is in just after the kick
(see Fig.\ref{fig:eng_str}).
\bibitem{note2}
In preparation for the simulation, the pendula are allowed to evolve 
for a sufficiently long time from an initial state
with total energy $E$. Then one pendulum is kicked,
instantaneously increasing its kinetic energy from nearly $0$ to $E_0$.
\end{thebibliography}
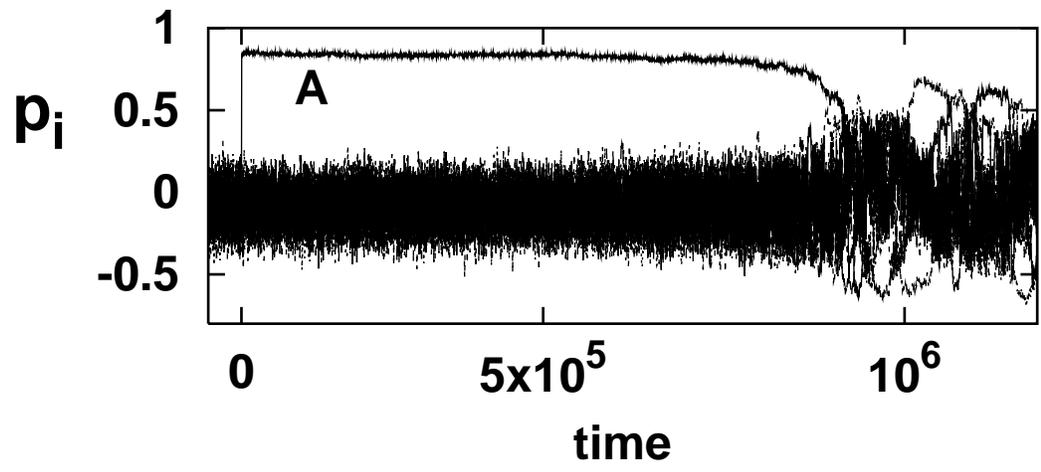
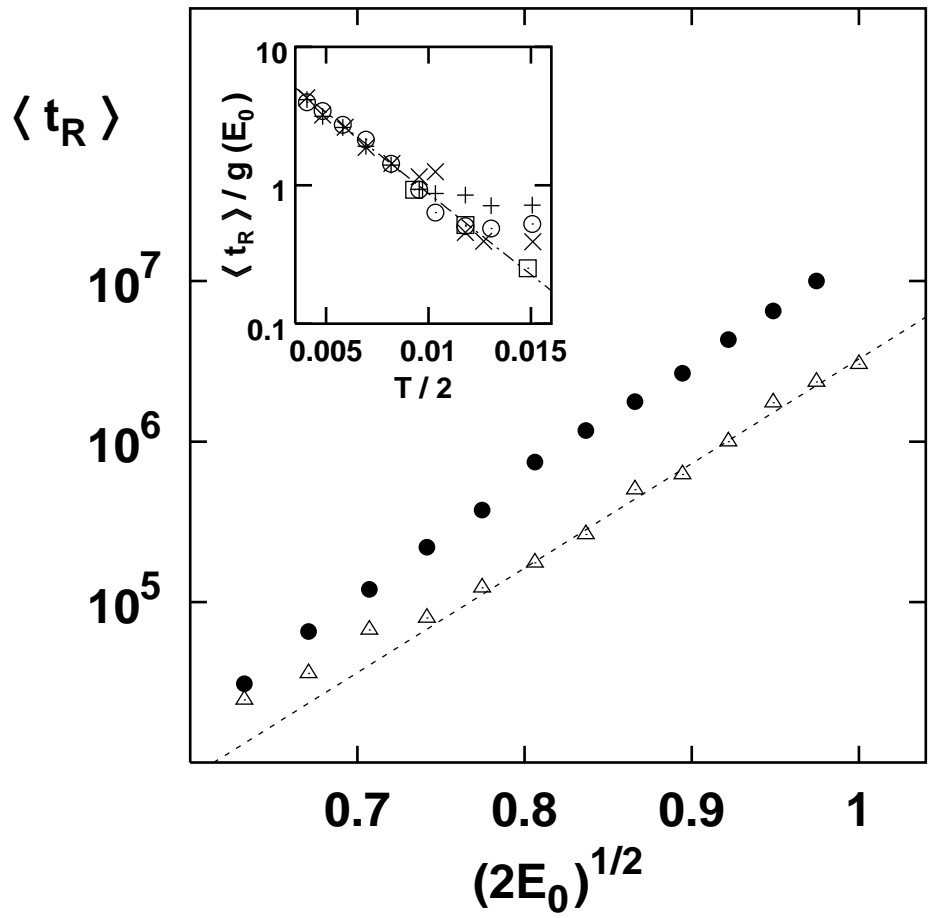
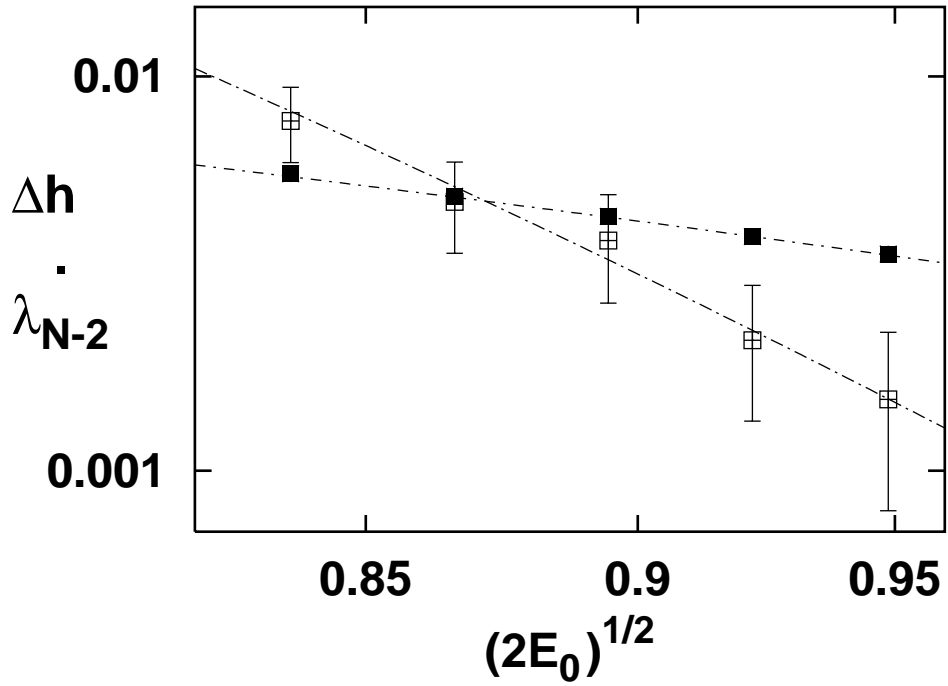
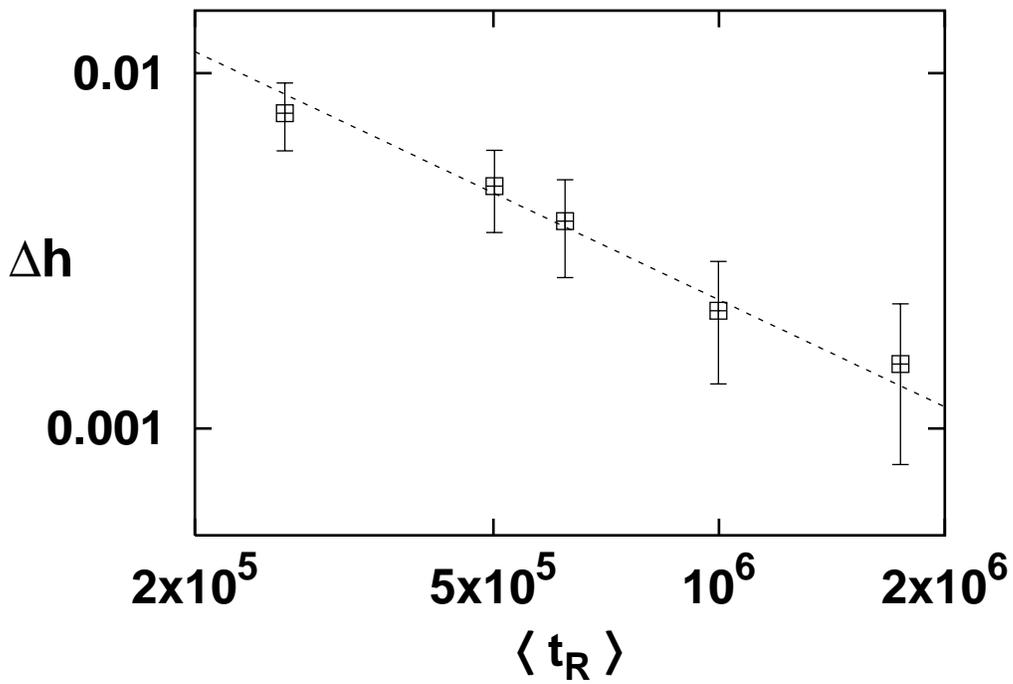
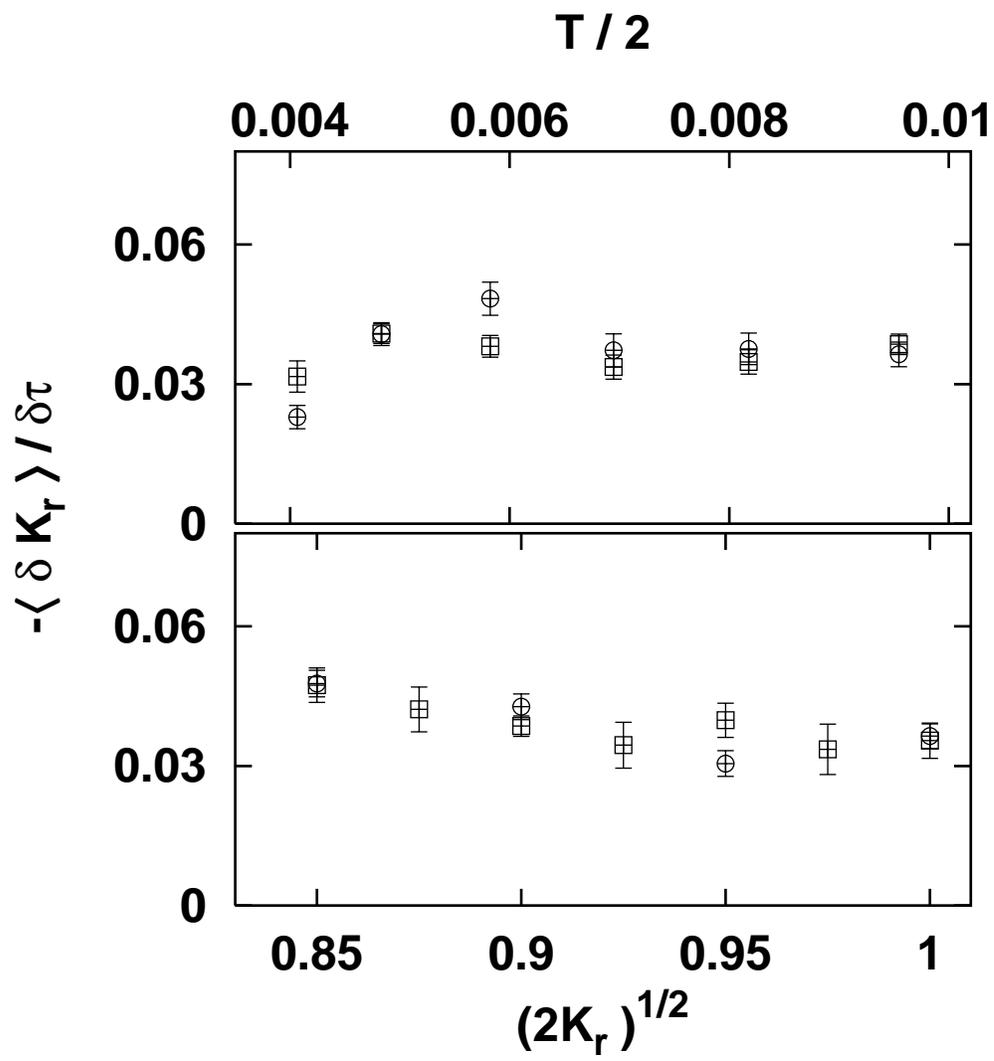
\end{document}